\documentclass[pra,twocolumn,showpacs,aps,superscriptaddress]{revtex4-1}
\usepackage{color}
\usepackage{amsmath,amssymb}
\usepackage{txfonts}
\usepackage{graphicx}

\usepackage[unicode]{hyperref}
\hypersetup{
    pdftitle={Vortex Decay},
    pdfauthor={S. J. Rooney, A. J. Allen, U. Z\"{u}licke, N. P. Proukakis and A. S. Bradley},
    pdfsubject={Decay of a superfluid vortex},
    colorlinks=true,
    linkcolor=blue,
    citecolor=blue,
    filecolor=black,
    urlcolor=blue
}
\usepackage{bm}
\usepackage{epstopdf}
\usepackage{enumerate}
\def\urlprefix{}
\def\url#1{}

\newcommand{\PP}{{\cal P}}
\newcommand{\QQ}{{\cal Q}}
\newcommand{\bra}[1]{\langle#1|}
\newcommand{\ket}[1]{|#1\rangle}
\newcommand{\ec}{\epsilon_{\textrm{cut}}}
\newcommand{\hpp}{\hat{\cal P}}
\newcommand{\hqq}{\hat{\cal Q}}
\newcommand{\ve}{\varepsilon}
\newcommand{\eref}[1]{(\ref{#1})} 
\newcommand{\fref}[1]{Fig.~\ref{#1}}

\newcommand{\rr}{\mathbf{r}}

\newcommand{\kk}{\mathbf{k}}

\newcommand{\EQ}[1]{\begin{align}#1\end{align}}

\newcommand{\rC}{C} 
\newcommand{\rI}{I} 

\newcommand{\LL}{L}

\newcommand{\gam}{\gamma}
\begin{document}
\title{Reservoir interactions of a vortex in a trapped 3D Bose-Einstein condensate}
\author{S.~J. Rooney} 
\affiliation{Department of Physics and Centre for Quantum Science, University of Otago, Dunedin 9010, New Zealand.}
\affiliation{Dodd-Walls Centre for Photonic and Quantum Technologies.}
\author{A.~J. Allen} 
\affiliation{Joint Quantum Centre (JQC) Durham-Newcastle, School of Mathematics and Statistics, Newcastle University, Newcastle upon Tyne NE1 7RU, England, UK}
\author{U. Z\"{u}licke} 
\affiliation{School of Chemical and Physical Sciences and MacDiarmid Institute for Advanced Materials and Nanotechnology, Victoria University of Wellington, P.O. Box 600, Wellington 6140, New Zealand.}
\affiliation{Dodd-Walls Centre for Photonic and Quantum Technologies.}
\author{N.~P. Proukakis} 
\affiliation{Joint Quantum Centre (JQC) Durham-Newcastle, School of Mathematics and Statistics, Newcastle University, Newcastle upon Tyne NE1 7RU, England, UK}
\author{A.~S. Bradley} 
\affiliation{Department of Physics and Centre for Quantum Science, University of Otago, Dunedin 9010, New Zealand.}
\affiliation{Dodd-Walls Centre for Photonic and Quantum Technologies.}
\begin{abstract}
We simulate the dissipative evolution of a vortex in a trapped finite-temperature dilute-gas Bose-Einstein condensate using first-principles open-systems theory. Simulations of the complete stochastic projected Gross-Pitaevskii equation for a partially condensed Bose gas containing a single quantum vortex show that the transfer of condensate energy to the incoherent thermal component without population transfer provides an important channel for vortex decay. For the lower temperatures considered, this effect is significantly larger that the population transfer process underpinning the standard theory of vortex decay, and is the dominant determinant of the vortex lifetime. A comparison with the Zaremba-Nikuni-Griffin kinetic (two-fluid) theory further elucidates the role of the particle transfer interaction, and suggests the need for experimental testing of reservoir interaction theory. The dominance of this particular energetic decay mechanism for this open quantum system should be testable with current experimental setups, and its observation would have broad implications for the dynamics of atomic matter waves and experimental studies of dissipative phenomena.
\end{abstract}
\maketitle
\section{Introduction} 
The dynamics of a quantum vortex in a Bose-Einstein condensate provides an observable and topologically stable probe of superfluid fluctuations~\cite{Pismen1999,Fetter2001,Thouless:2007gl,Fetter09a,Rooney:2010dp,Rooney:2011fm,Rooney:2013ff}, and thus a test of dynamical theories of finite-temperature Bose-Einstein condensates~\cite{Blakie:2008isa,Proukakis:2008eo}. The precise nature of vortex dynamics underpins emergent behavior in quantum turbulence~\cite{Barenghi:2014gj,Neely:2013ef,Rooney:2013ff,Kwon:2014ud,Kwon:2015ch}, and is of increasing relevance in theories of spin-orbit coupled~\cite{Fetter:2014kra} and spinor~\cite{Kawaguchi:2012bl,Seo:2015ho} condensates. While Hamiltonian vortex motion is well understood at the mean field level, dissipation plays a central role in the creation of spontaneous vortices~\cite{Weiler:2008eu} and solitons~\cite{Lamporesi:2013bi} during the BEC phase transition, in the formation of negative temperature states~\cite{Henn09a,Neely:2013ef,Reeves:2013hy,Billam:2014hc,Simula:2014ku}, in the formation~\cite{Rooney:2013ff} and break-down~\cite{Eckel:2014gf} of persistent currents, and the frustrated equilibration of spinor condensates~\cite{Liu:2009en,Bradley:2014a}. A variety of theoretical techniques have been used to study finite-temperature vortex dynamics \cite{Fedichev1999,Berloff:2014cu}, including phenomenological damping of the Gross-Pitaevskii equation~\cite{Tsubota2002,Penckwitt2002,Madarassy:2008in}, two-fluid models~\cite{Kobayashi:2006kb,Jackson:2009jo,Allen:2013cs}, the projected Gross-Pitaevskii equation~\cite{Wright:2008ha,Wright:2009eh,Wright:2010pj} and related classical field theories~\cite{Schmidt2003,Shukla:2013kv}, and the stochastic Gross-Pitaevskii equation~\cite{Duine2004,Bradley:2008gq,Rooney:2010dp,Rooney:2011fm}. However, the dissipative motion due to reservoir interactions of a quantum vortex have yet to be tested against experimental observations~\cite{Anderson2000,Anderson:2010bd,Freilich10a,Middelkamp:2011fq,Navarro:2013hb}.
\begin{figure}[!t]
\begin{center}
\includegraphics[width=0.8\columnwidth]{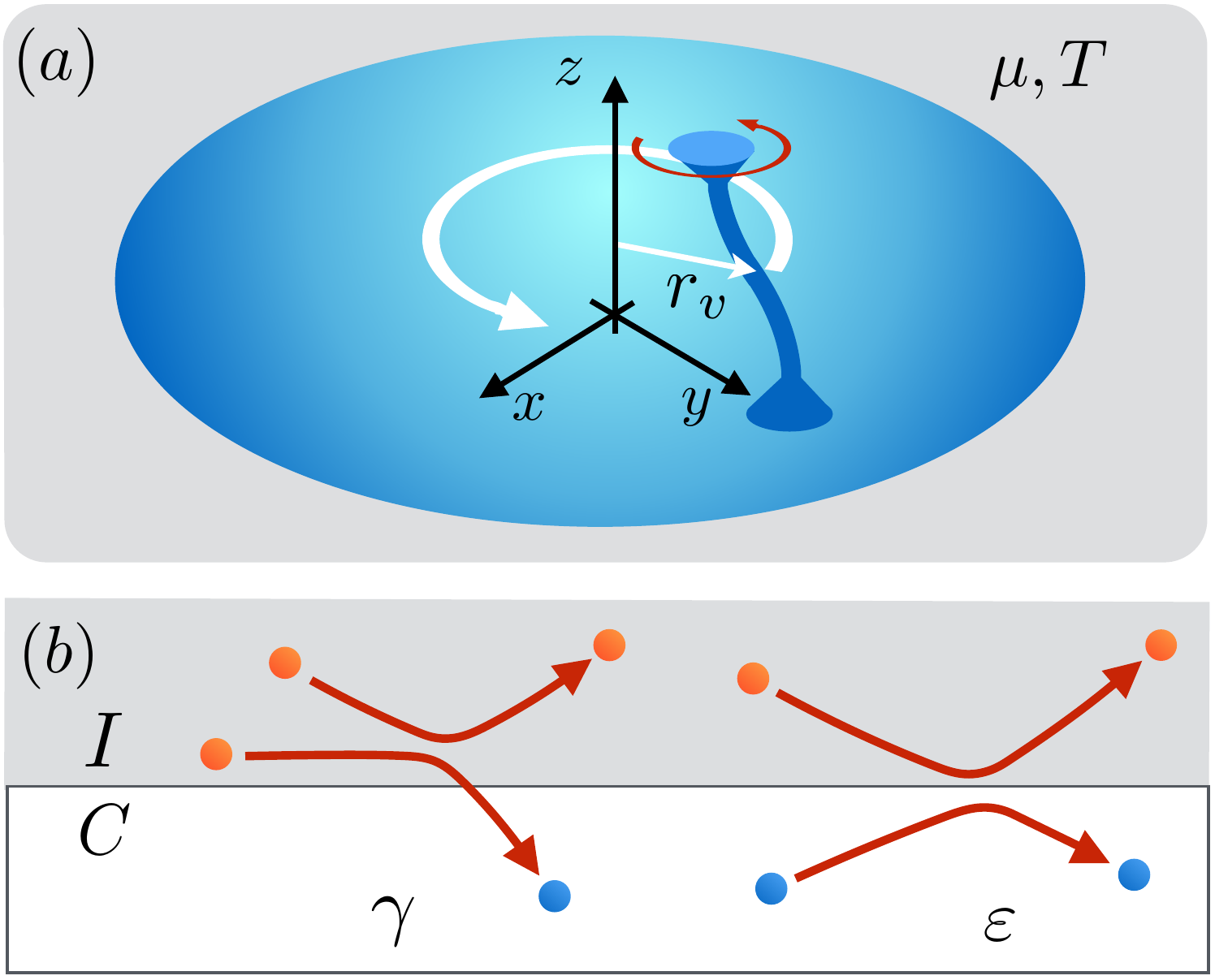}
\caption{(Color online) (a) Schematic of a quantum vortex in an oblate BEC, undergoing precession around the trap symmetry axis ($z$). The trap anisotropy causes vortices to align with the $z$ axis and partially suppresses vortex bending modes. (b) Reservoir interaction processes in the SPGPE high-temperature reservoir theory of the trapped Bose gas.
We distinguish 2 different processes between coherent ($\rC$) and incoherent ($\rI$) regions: collisional particle-transfer processes, labelled by $\gamma$, and energy-damping processes which involve energy transfer without particle transfer, labelled here by
$\varepsilon$.
}
\label{fig1}
\end{center}
\end{figure}
\par
In this work we perform first-principles, \emph{ab initio} simulations of a realistic experimental system, as a test of high-temperature open systems theory for Bose gases. We use the stochastic projected Gross-Pitaevskii equation (SPGPE)~\cite{Gardiner:2003bk,Rooney:2012gb} to model the decay of a lone quantum vortex in an oblately confined Bose-Einstein condensate at appreciable temperature. As a central vortex is only metastable, dissipative processes cause the vortex to exit the condensate, thus removing its angular momentum. Typically, the dominant reservoir interaction involves the transfer of particles between the reservoir and the condensate,  
with this process essential for modelling condensate growth from a quenched vapor~\cite{Bradley:2008gq,Weiler:2008eu,Su:2013dh}. An additional reservoir interaction included in the full SPGPE theory~\cite{Gardiner:2003bk,Rooney:2012gb} involves number-conserving transfer of energy between the reservoir and the condensate. While  
the first process plays a central role in most dissipative BEC theory~\cite{Zaremba:1999iu,Tsubota2002}, and often provides a reasonable qualitative picture of the important damping processes, 
the latter process appears to be increasingly important for quantitive modelling of dissipative phenomena; indeed, there may be systems where 
such a damping mechanism becomes the dominant reservoir interaction. In this work we find that vortex decay provides an example of such a system. Of the two reservoir interaction processes contained within the full SPGPE, we find that 
energy-damping without population transfer between system and reservoir provides the dominant contribution to the vortex decay rate; quantitatively,  
such terms reduce the lifetime by up to a factor of 4 compared to the predictions of the `simple growth' SPGPE~\cite{Bradley:2008gq}. Comparison of single trajectories of our lowest temperature SPGPE simulations with the two-fluid ZNG theory~\cite{Zaremba:1999iu,Jackson:2009jo} shows qualitative agreement with the simple growth SPGPE, emphasising the significance of vortex decay as a test of the  
dominant damping process in finite temperature BEC theory. Furthermore, the full SPGPE results for vortex lifetimes are quantitatively accounted for by simulating the reduced SPGPE containing \emph{only} energy-damping reservoir interactions.

\section{System and Theory}
\subsection{System}
We consider a dilute gas of bosonic atoms (with mass $m$) interacting via s-wave collisions (scattering length $a$), described by the Hamiltonian
\EQ{\label{Hdef}
H&=\int d\rr\;\hat\Psi^\dag(\rr)\left[{\cal H}(\rr)+\frac{g}{2}\hat\Psi^\dag(\rr)\hat\Psi(\rr)\right]\hat\Psi(\rr),
}
where
\EQ{\label{Hdens}
{\cal H}(\rr)&=-\frac{\hbar^2\nabla^2}{2m}+V_0(\rr),
}
and the confining potential
\EQ{\label{vdef}
V_0(\rr)&=\frac{m}{2}\left[\omega_\perp^2(x^2+y^2)+\omega_z^2 z^2 \right],
}
is parabolic with cylindrical symmetry. In the cold-collision regime the interaction parameter is given by $g=4\pi\hbar^2 a/m$. The field operators obey the standard commutation relations for bosons, with non-vanishing commutator
\EQ{\label{bcom}
[\hat\Psi(\rr),\hat\Psi^\dag(\rr')]&=\delta(\rr-\rr')
}
We consider a fixed total particle number $N$, and a range of temperatures $T\lesssim T_c(N)$, appreciable compared to the ideal gas transition temperature 
\EQ{\label{Tc}
T_c(N)&=\frac{\hbar\bar\omega}{k_B} \left(\frac{N}{\zeta(3)}\right)^{1/3},
}
for geometric mean frequency $\bar\omega^3=\omega_z\omega_\perp^2$. The condensate is assumed to initially contain a single quantum vortex coincident with the $z$-axis, and due to the temperature, is immersed in a thermal cloud. The non-condensate fraction is assumed non-rotating (as will typically be the case unless the trap is made extremely symmetric), so that the initial central vortex, contributing angular momentum of $\hbar$ per condensate particle, is out of equilibrium with the thermal cloud. However, a central vortex is a thermodynamic metastable state~\cite{Fetter2001}, and due to the trap symmetry the instability must be initiated by a symmetry breaking perturbation.
An off-centre vortex is thermodynamically unstable, and thermal fluctuations perturb the vortex away from the otherwise Hamiltonian precession of the vortex on closed (angular momentum conserving) circular paths around the $z$-axis~\cite{Fetter2010a} shown schematically in \fref{fig1} (a). 
\subsection{SPGPE Theory}
The SPGPE is the equation of motion of a classical field describing the low-energy coherent ($C$) region of matter waves, in contact with a reservoir consisting of a high-energy incoherent ($I$) region. The two regions are identified by introducing an energy cutoff, $\ec$, and an associated orthogonal projection operator that enforces the cutoff rigorously. Carrying out the derivation leads to a Gross-Pitaevskii-like equation, with additional damping and noise terms~\cite{Gardiner:2003bk,Bradley:2014a}. The dissipative interactions described by the SPGPE are: (i) 
processes involving Bose-enhanced collisions between $I$-region and $C$-region particles, which result in population and energy transfer between the two subsystems, as well as (ii) 
processes transferring energy between the $I$ and $C$ regions, without associated particle transfer; the two processes are shown schematically in \fref{fig1} (b). Despite extensive use of the damped GPE obtained as the low-temperature limit of (i) to model dissipative phenomenology~\cite{Choi:1998eh,Tsubota2002,Penckwitt2002,Neely:2013ef,Reeves:2013hy}, rigorous experimental tests of dissipative dynamics that can distinguish between (i) and (ii) are currently lacking.
\par
In SPGPE theory, the system is represented in terms of a set of single-particle eigenfunctions $\phi_n(\rr)$, that satisfy
\EQ{\label{speigs}
{\cal H}(\rr)\phi_n(\rr)=\epsilon_n\phi_n(\rr),
}
where $n$ denotes all quantum numbers required to specify a unique eigenstate. In terms of these eigenfunctions, the field operator is
\EQ{\label{fop}
\hat\Psi(\rr)&=\sum_n\hat a_n\phi_n(\rr),
}
for single-mode operators $\hat a_n$ satisfying $[\hat a_n,\hat a_m^\dag]=\delta_{nm}$.
Our first task is to consistently separate the system into a $C$-region, where populations are appreciable and the atoms are at least partially coherent, and an $I$-region where populations are low and atoms are incoherent. We then seek an equation of motion for the $C$-region, treating the $I$-region as an incoherent reservoir. A significant advantage of effecting the separation in the single-particle basis is that at sufficiently high energy it diagonalises the many-body problem, thus providing a good basis for separating the system. We define the $C$-region as $C=\{\epsilon_n\leq\epsilon_{\textrm{cut}}\}$, where $\ec$ will be significantly larger than the system chemical potential $\mu$ (of order $2\mu$ to $3\mu$). In this basis we introduce the orthogonal projection operators 
\EQ{\label{Pdef}
\hpp&\equiv\sum_{\epsilon_n\leq\ec}\ket{n}\bra{n},\\
\hqq&\equiv 1-\hpp,
}
satisfying $\hpp\hpp=\hpp$, $\hqq\hqq=\hqq$, $\hqq\hpp=0$. In the position representation the field operator decomposes into $\rC$-region and $I$-region operators as
\EQ{\label{cfield}
\hat\Psi(\rr)&= \PP\hat\Psi(\rr)+\QQ\hat\Psi(\rr)\equiv \hat\psi(\rr)+\hat\eta(\rr)
}
respectively, where 
\EQ{\label{Px}
\hat\psi(\rr)&= \PP\hat\Psi(\rr)\equiv\sum_{\epsilon_n\leq \ec}\phi_n(\rr)\int d^3\rr\;\phi^*_n(\rr)\hat \Psi(\rr)\\
&=\sum_{\epsilon_n\leq \ec}\hat a_n\phi_n(\rr),
}
defines the spatial representation of $\hpp$ and the projected field operator with commutator
\EQ{\label{dC}
[\hat\psi(\rr),\hat\psi^\dag(\rr')]=\delta(\rr,\rr')\equiv \sum_{\epsilon_n\leq \ec}\phi_n(\rr)\phi^*_n(\rr').
}
This formal separation of the system provides a natural approach to deriving an equation of motion describing the evolution of the $\rC$-region, the details of which can be found elsewhere~\cite{Gardiner:2003bk,Bradley:2014a}. The derivation proceeds by mapping the master equation for the $\rC$-region density operator to an equation of motion for the Wigner distribution of the system. Truncating the third order (super-diffusive) terms that appear in this generalized Fokker-Planck equation (FPE) leads to an FPE containing only drift and diffusion terms. This equation of motion for the Wigner distribution may then be mapped to a stochastic differential equation for a classical field $\psi(\rr)$ (the $C$-field), the moments of which correspond to symmetrically ordered averages of the field operator at equal times~\cite{Steel:1998jr,Blakie:2008isa}. 
\par
For our purposes, we take as our starting point the three-dimensional Stratonovich SPGPE for the $\rC$-field~\cite{Rooney:2012gb}. Taking our energy reference as $\mu$, the SPGPE takes the form
\EQ{\label{3dSPGPE}
(S) d\psi(\rr,t)&=d\psi\Big|_H+d\psi\Big|_\gamma+(S)d\psi\Big|_\ve,
}
with
\begin{subequations}
\label{spgpeAll}
\EQ{
\label{spgpeH}
i\hbar d\psi\Big|_H&=\PP\left\{\LL\psi dt\right\},\\
\label{spgpeG}
i\hbar d\psi\Big|_\gamma&=\PP\left\{-i\gamma\LL\psi dt+i\hbar dW(\rr,t)\right\},\\
\label{spgpeE}
(S)i\hbar d\psi\Big|_\ve&=\PP\left\{U^\ve(\rr,t)\psi dt-\hbar\psi dU(\rr,t)\right\}.
}
\end{subequations}
The Hamiltonian evolution described by \eref{spgpeH} is generated by the nonlinear operator 
\EQ{\label{Ldef}
\LL\psi(\rr,t)&\equiv \left[{\cal H}(\rr)+g|\psi(\rr,t)|^2-\mu\right]\psi(\rr,t),
}
thus recovering the PGPE~\cite{Blakie05a}
\EQ{\label{pgpe}
i\hbar\frac{\partial \psi(\rr,t)}{\partial t}\Big|_H&=\PP\{ L\psi(\rr,t)\},
}
describing the evolution of a low-energy fraction of atoms with partial coherence, including the condensate and a band of low-energy excitations~\cite{Davis2001a,Blakie05a,Wright:2011ey}.
The $\rI$-region reservoir coupled to the PGPE resides at energies above $\ec$, and is described by a semi-classical Bose-Einstein distribution with chemical potential $\mu$, temperature $T$. The variables $\mu, T, \ec$ play a central role in setting the strength of reservoir interaction processes~\cite{Bradley:2008gq,Rooney:2012gb,Bradley:2014a}. 

Eq.~(\ref{spgpeG}) contains the coupling between the two subsystems associated with transfer of particles between them, as illustrated in Fig.~1(b) [left plot]. In previous literature, the addition of this term to \eref{spgpeH} is often called the \emph{simple growth} SPGPE~\cite{Bradley:2008gq}. Here we adopt the convention of \cite{Bradley:2015cx} and refer to the equation of motion as the $\gamma$-SPGPE. Without formal projection, the $\gamma$-SPGPE becomes an equation that is referred to as the SGPE, and is very similar in spirit and detail to that derived by Stoof~\cite{Stoof:1999tz,Stoof:2001wk,Proukakis:2008eo}.

Eq.~(\ref{spgpeE}) represents an important additional term, originally referred to as the \emph{scattering term} (and called the $\varepsilon$-SPGPE in \cite{Bradley:2008gq}), which leads to energy transfer between the two subsystems, without any associated population transfer. We will refer to the two processes arising from Eqs.~(\ref{spgpeG}), (\ref{spgpeE}) as \emph{number-damping} and \emph{energy-damping} (although the latter term is not the only mechanism which damps energy from the $C$ region). The equations of motion shall henceforth be termed the full SPGPE
\eref{3dSPGPE}, the $\gamma$-SPGPE [\eref{spgpeH}+\eref{spgpeG}], and the $\varepsilon$-SPGPE [\eref{spgpeH}+\eref{spgpeE}].

Within the SPGPE theory, the interaction with the $C$-region can be cast in terms of the functions~\cite{Gardiner:2003bk,Bradley:2008gq,Rooney:2012gb}
\begin{subequations}
\label{spgpeParams}
\EQ{\label{gamdef}
\gamma(\mu,T,\ec)&=\frac{8 a^2}{\lambda_{dB}^2}\sum_{j=1}^\infty \frac{e^{\beta\mu(j+1)}}{e^{2\beta\ec j}}\Phi\left[\frac{e^{\beta\mu}}{e^{2\beta\ec}},1,j\right]^2,\\
U^\ve(\rr,t)&=-\hbar \int d^3\rr^\prime \ve(\rr-\rr^\prime)\nabla^\prime\cdot\mathbf{j}(\rr^\prime,t),\\
\mathbf{j}(\rr,t)&=\frac{i\hbar}{2m}\left[\psi\nabla\psi^*-\psi^*\nabla\psi\right],\\
\ve(\rr)&=\frac{\cal M}{(2\pi)^3}\int d^3\kk \frac{e^{i\kk\cdot \rr}}{|\kk|},\\
{\cal M}(\mu,T,\ec)&= \frac{16 \pi a^2}{e^{\beta(\ec-\mu)}-1},\label{mdef}
}
\end{subequations}
where $\beta=1/k_B T$, $\lambda_{dB}=\sqrt{2\pi \hbar^2/m k_B T}$ is the thermal de Broglie wavelength, and $\Phi[z,x,a]=\sum_{k=0}^\infty z^k/(a+k)^x$ is the Lerch transcendent, and where the reservoir coupling rates $\gamma$ and $\ve(\rr)$ are both dimensionless. The rates \eref{gamdef}, \eref{mdef} are found by analytically evaluating the relevant collision integrals over all $I$-region particles that contribute to the interactions shown in \fref{fig1} (b); rigorous restriction of the integrals to the phase space of the $I$-region generates the cutoff dependence of these rates, and is crucial for setting consistent damping parameters in SPGPE simulations~\cite{Bradley:2008gq,Rooney:2012gb}. We emphasize that in this formulation $\mu$ is also a function of $T,\ec$, and $N$, being found self-consistently for a given system temperature and atom number~\cite{Rooney:2010dp}. Furthermore, the position independent form of $\gamma$ in Eq.~\eref{gamdef} is a consequence evaluating the collision integrals analytically when $\mu\ll\ec$ (neglecting the weak position dependence near the $C$-region boundary)~\cite{Bradley:2008gq}. 
\par
The noise terms are Gaussian, with non-vanishing correlations
\begin{subequations}
\label{allNoise}
\EQ{
\langle dW^*(\rr,t)dW (\rr^\prime,t)\rangle&=\frac{2k_B T }{\hbar}\gamma\delta(\rr',\rr) dt, \\
\langle dU(\rr,t)dU(\rr^\prime,t)\rangle&=\frac{2k_B T}{\hbar}\ve(\rr-\rr^\prime) dt.
}
\end{subequations}
The form of \eref{allNoise} constrains the equilibrium solutions of \eref{spgpeAll} to automatically satisfy the fluctuation-dissipation theorem. We emphasize that the noise in \eref{spgpeG} is complex, while the noise in \eref{spgpeE} is real; the former thus provides a source of particles, while the latter may be interpreted as a stochastic potential. 
\par
In deriving the equation of motion~\eref{3dSPGPE}, there are some approximations that should be noted. Firstly, the complete generalized FPE for the Wigner distribution contains third order derivatives with respect to the fields, arising from the two-body interaction term. Assuming significant population per mode, these terms may be safely neglected~\cite{Graham:1973te}, and this truncated Wigner approximation (TWA) has been used extensively in treatments of ultra-cold Bose gases~\cite{Steel:1998jr,Blakie:2008isa,Polkovnikov10a}. Secondly, an energy cutoff has been imposed in terms of single particle modes~\cite{Davis2001b}. While numerically and formally convenient, the basis is not strictly an exact one for implementing an energy cutoff for the many-body system. However, at sufficiently high energies, the interacting many-body system is diagonal in the basis of single particle states --- a property well known in Bogoliubov theory. The satisfaction of both constraints is the reason for choosing $\ec \sim 2\mu-3\mu$ in $C$-field theory. At low temperatures the requirement that the mode population is of order 1 at the cutoff energy conflicts with the requirement that the cutoff is large.
The two conditions introduce competing constraints, and validity of the approximations becomes questionable. For the lowest temperature considered in this work, $T=0.6T_c$, we are working at the edge of the validity regime, with cutoff $\ec=1.7\mu$ [see Table \ref{paramsMod}].
\par
\renewcommand{\tabcolsep}{7pt}
\begin{table}[!h]
\begin{center}
    \begin{tabular}{ lccccc}
    \hline\hline
    $T/T_c(N)$ & $0.6 $ & $0.65 $ & $0.7 $ & $0.75 $ & $0.8 $ \\ \hline 
    $T$ [{\rm nK}]& $278\;  $ & $301 $ & $324 $ & $347  $ & $370 $ \\ 
$\mu/\hbar\bar\omega$  & $15.54$ & $14.65$ & $13.61$ & $12.37$ & $10.09$\\ 
   $\epsilon_{\rm cut}/\hbar\bar\omega$ &  $26.35$ & $27.80$ &$29.07$ & $30.13$ & $30.93$\\ 
   $Z$ & $2970$ & $3663$ &$4180$ & $4456$ & $5044$\\
    $\gamma \times 10^3$ &  $2.1$ & $1.8$ & $1.7$ & $1.6$ & $1.6$\\ 
    ${\cal M}  \frac{k_B T x_0^2}{\hbar}\times 10^3$ &  $5.1 $ & $4.5$ & $4.0$ & $3.6$ & $3.4$\\ 
    $r_f \;[\mu m]$ & 4.02 & 3.90 & 3.76 & 3.59 & 3.36 \\\hline \hline
    \end{tabular}
      \caption{Parameters used for the SPGPE simulations, for a range of $T$ values. At fixed $N=8\times 10^4$ atoms of $^{87}{\rm Rb}$, the critical temperature is $T_c(N)=463 {\rm nK}$. $Z$ is the total number of single-particle modes in the $C$ region. The final row gives the spatial boundary of the region in which the vortex is detected, with cutoff radius $r_f=0.65R_{\rm TF}$ depending on the system temperature (see text).}
         \label{paramsMod}
\end{center}
\end{table}

\section{SPGPE Simulations} 
\subsection{System parameters, modelling, and observables} 
The specific system we model contains a total of
$N=8\times 10^4$ $^{87}{\rm Rb}$ atoms, with trap frequencies $(\omega_\perp,\omega_z)=2\pi\times (150,600)$, producing an oblate system with $\omega_z/\omega_\perp=4$. The s-wave scattering length is $a_s = 100a_0$, where $a_0$ is the Bohr radius. We consider the dynamics of the distance of the vortex from the $z$-axis, $r_v(t)$, as a function of system temperature for $0.6\leq T/T_c(N)\leq 0.8$, where $T_c(N)=463 {\rm nK}$. The oblate geometry has two functions. Firstly, it causes axial alignment of the vortex (assisting optical imaging in experiments~\cite{Neely:2013ef}), and secondly, it suppresses bending (Kelvin) modes caused by thermal fluctuations~\cite{Rooney:2011fm}, giving approximately 2D vortex dynamics. Our choice of relatively weak oblateness means that the condensate is far from the quasi-condensate regime and retains global phase coherence. This choice also allows significant bending of the vortex line, thus increasing coupling to reservoir modes and accelerating vortex decay~\cite{Rooney:2011fm}.
\par
Finding consistent SPGPE parameters forms an essential part of the ab initio modelling process~\cite{Rooney:2013ff}. For a given $T/T_c$, we find SPGPE parameters so that the total number of atoms (contained in both $\rI$- and $\rC$-regions) is $N\simeq 8\times10^4$, and so that the chosen cutoff $\ec$ is consistent with the validity condition for classical field theory, namely, that the mean thermal occupation of the $I$-region modes is at most of order unity. We then use the Penrose-Onsager criterion to determine the condensate number $N_0$ of our equilibrium ensembles, checking that the condensate fraction is consistent with $T/T_c$. 
\par
Our SPGPE simulations use a cutoff that includes a total of $Z$ single-particle modes in the $\rC$-field region where $Z$ is of order $10^3$, as shown in Table \ref{paramsMod}. We calculate ensembles of $100$ trajectories for each temperature and sub-theory, integrating using either the Runge-Kutta method for $\gamma$-SPGPE or the semi-implicit method for $\ve$-SPGPE and full SPGPE. Details of the numerical integration algorithms are given in Ref.~\cite{Rooney:2014kc}. The main observable calculated in this work is the the distance of the vortex from the $z$ axis at $z=0$, $r_v(t)\equiv \sqrt{x_v(t)^2+y_v(t)^2}$. We extract this quantity from each trajectory, and compute the ensemble average $\langle r_v(t)\rangle$. 
\begin{figure}[!t]
\begin{center}
\includegraphics[width=\columnwidth]{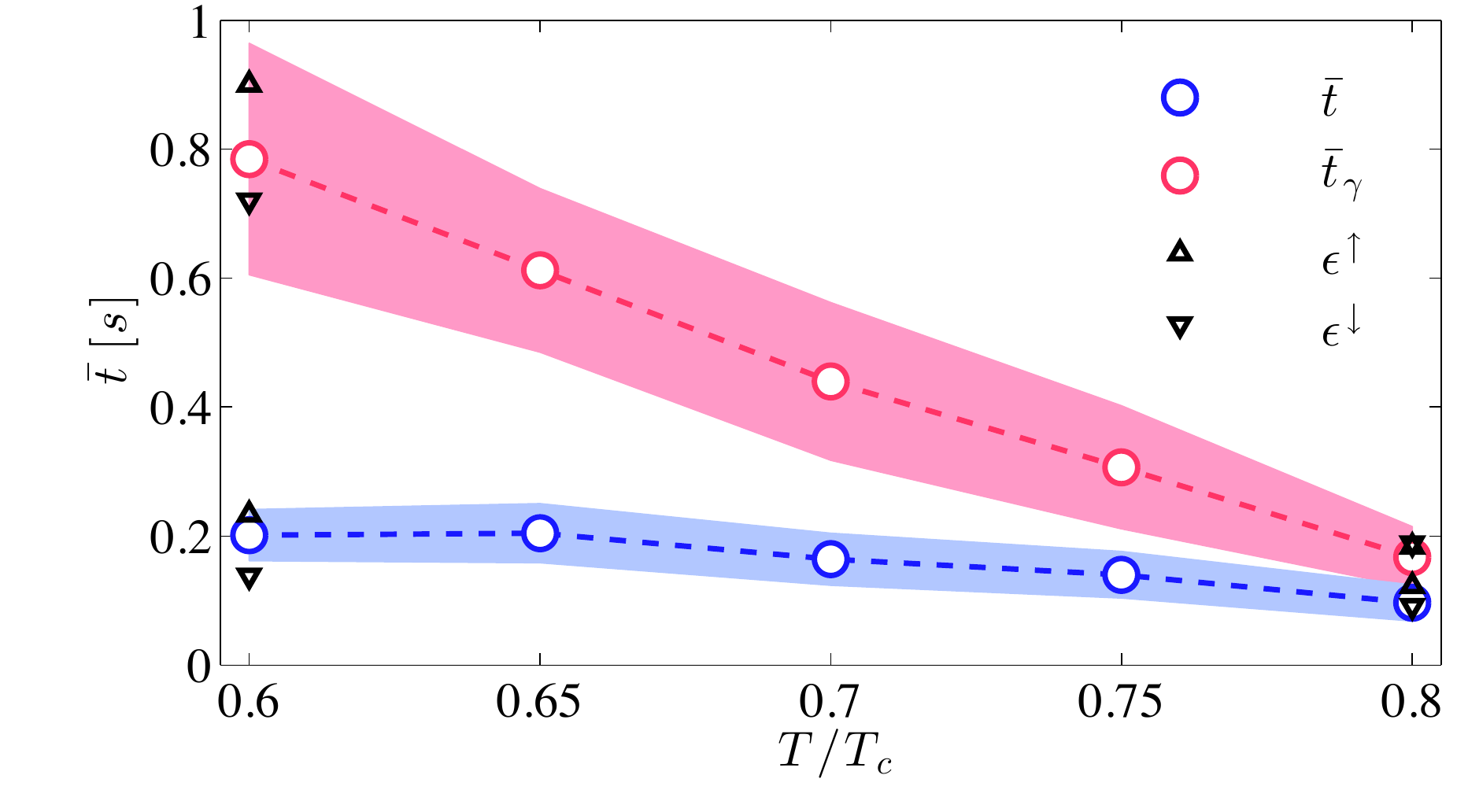}
\caption{(Color online) Vortex lifetime as a function of relative temperature, for the full SPGPE (blue) and the $\gamma$-SPGPE (red). The markers show the ensemble-averaged lifetime $\bar t$. The spread of observable lifetimes is indicated by the shaded region showing one standard deviation of the ensemble of single trajectory lifetimes; the mean lifetime has converged for our ensemble of $100$ trajectories at each temperature. The effect of consistently varying the cutoff $\ec$ on the vortex lifetime is shown for the temperatures $T=0.6T_c$ and $T=0.8T_c$ (See Section \ref{sec:cutoff}). }
\label{fig:lifetime}
\end{center}
\end{figure}

\subsection{Temperature dependence of SPGPE vortex lifetimes }
We initialize the vortex at $r_v(0)\equiv r_0=0$, and carry out a systematic study of vortex lifetimes as a function of $T/T_c$, over a regime where the SPGPE is valid, and where the vortex should be clearly visible in column density imaging~\cite{Rooney:wi}. Initializing the vortex in the $C$-field for a thermal equilibrium non-rotating BEC proceeds as in Ref.~\cite{Rooney:2010dp}. First, a vortex-free state is obtained by integrating the SPGPE from a Thomas-Fermi initial state, until all transients have decayed and the $C$-field reaches equilibrium at the required $T, \mu(N,T)$. Then a central vortex is phase imprinted onto the complete $C$-field. As this procedure is phase-coherent, the net effect is that only the condensate fraction is set into rotation, while the imprinted phase has very little effect on the non-condensate part of the $C$-field. This creates a consistent $C$-field for a non-rotating finite-temperature BEC containing a central, axially aligned vortex~\footnote{Reservoir interactions with the non-rotating $I$-region cause any small rotation imparted to the non-condensate part of the $C$-field to dissipate rapidly under time evolution.}. The ensemble of central vortex states is then used to initialize an ensemble of SPGPE trajectories computed as distinct numerical solutions of the stochastic differential equation~\eref{3dSPGPE}, and its variants.
\par
\fref{fig:lifetime} compares the mean vortex lifetime, $\bar{t}$, computed as the ensemble-average of the vortex lifetime found for each trajectory of the ensemble. The trajectory lifetime is extracted numerically as the time $t_l$ for the vortex to decay from $r_v(0)= 0$ to $r_v(t_l)=r_f=0.65 R_{\rm TF}$. We estimate the  TF-radius for the condensate as $R_{\rm TF}=\sqrt{2\mu( T,N,\ec)/m\omega_r^2}$, and so $r_f$ decreases with increasing temperature as shown in Table \ref{paramsMod}. Note that for the temperature range considered, the vortex is underdamped, executing a minimum of 5 precessional orbits (and up to $\sim 20$ at the lowest temperature) during decay in each trajectory. As may be clearly see in \fref{fig:lifetime}, the full SPGPE causes the vortex to decay significantly faster than the $\gamma$-SPGPE, with the difference becoming increasingly important at lower temperatures. This large difference between $\gamma$-SPGPE and full SPGPE (up to a factor of 4 in $\bar t$) is the main result of our work, and should be testable in experiments. We emphasize that the data in \fref{fig:lifetime} are obtained from first-principles $C$-field theory and require no fitted parameters. 
\par
\begin{figure*}[!t]
\begin{center}
\includegraphics[width=0.88\textwidth]{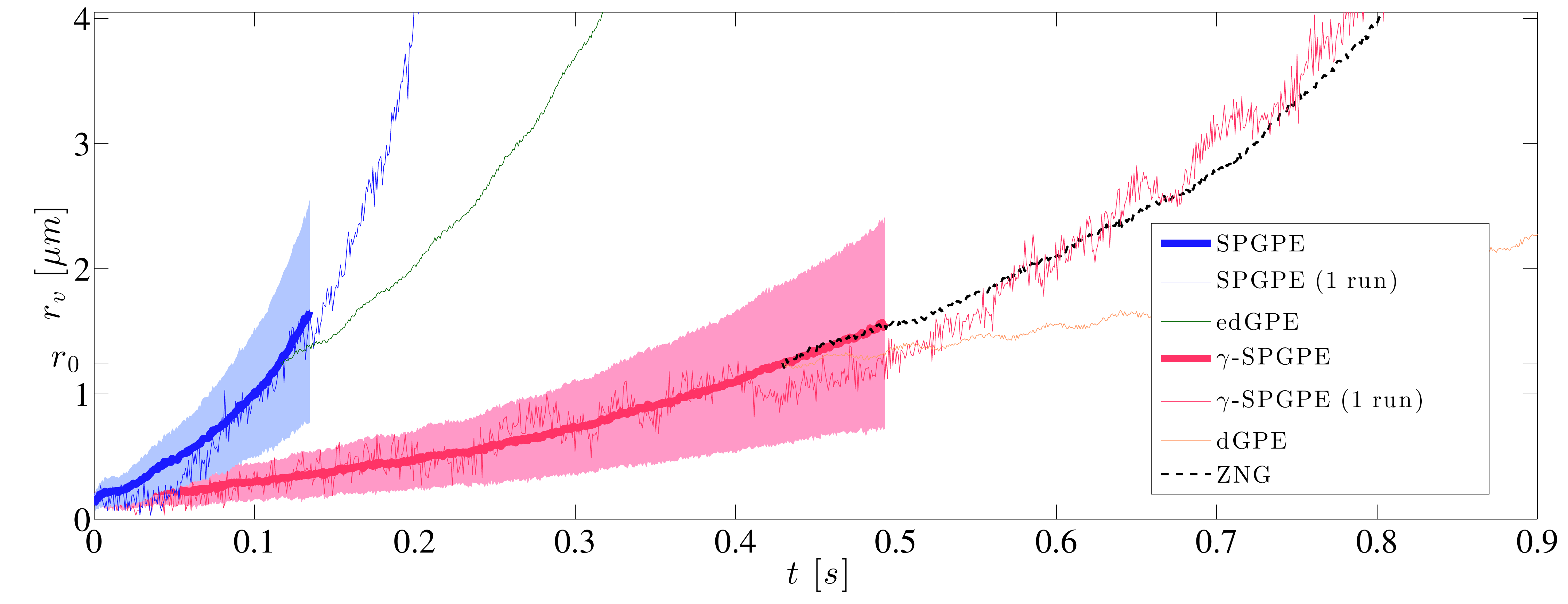}
\caption{(Color online) Distance of the vortex from the trap axis, $r_v(t)$, for ensemble averages (thick lines) and single trajectories. Ensemble averages are only computed up to the time of the first vortex exiting the detection region $r_v(t)\leq r_f=0.65R_{TF}=4.02\mu{\rm m} $ [See Table \ref{paramsMod}]. Also shown are the vortex dynamics for damped equations of motion corresponding to each stochastic theory (see text), starting from $r_v(0)=r_0\equiv 1.3\mu{\rm m}$.  The shaded regions show the range of lifetimes corresponding to one standard deviation of the ensemble.}
\label{fig1b}
\end{center}
\end{figure*}
The convergence of $\gamma$-SPGPE and full SPGPE as $T\to T_c$, evident in \fref{fig:lifetime}, is broadly consistent with earlier work on vortex decay using the $\gamma$-SPGPE~\cite{Rooney:2010dp,Rooney:2013ff}. The limiting behavior seen in Ref.~\cite{Rooney:2010dp} as $T\to T_c$ involved the vortex lifetime in $\gamma$-SPGPE converging to a value comparable to that predicted by the projected Gross-Pitaevskii equation, i.e. the Hamiltonian equation of motion for the $C$-field obtained by neglecting coupling to $I$. This limiting behavior is expected as low-energy thermal fluctuations in the $C$-field are dominant near the critical point~\cite{Davis:2006ic}. In Ref.~\cite{Rooney:2013ff} the dynamics of persistent current formation at high temperature ($T\sim 0.9T_c$) was described very accurately by the $\gamma$-SPGPE, also consistent with the present work. The increasing departure of the full SPGPE from $\gamma$-SPGPE at low temperatures is also consistent with the energy and number characteristics of vortex decay: the loss of a single vortex from a large BEC does not greatly alter the condensate population, but causes an appreciable change to the energy per particle~\cite{Fetter2001}. 
\subsection{Cutoff independence}\label{sec:cutoff}
\renewcommand{\tabcolsep}{13pt}
\begin{table}[!t]
\begin{center}
    \begin{tabular}{ lcccc}
    \hline\hline
    Parameters & $\epsilon^{\downarrow}_{\rm cut}$ & $\epsilon^{\uparrow}_{\rm cut}$ & $\epsilon^{\downarrow}_{\rm cut}$ & $\epsilon^{\uparrow}_{\rm cut}$ \\ \hline
    $T/T_c$ & 0.6 & 0.6 & 0.8 & 0.8 \\ 
$\gamma/\gamma_0$  & $1.52$ & $0.66$ & 1.31 & 0.75\\ 
   ${\cal M}/{\cal M}_0$ &  $1.41$ & $0.76$ & 1.24 & 0.82\\ 
    $n_{\rm cut}$ &  $1.21$ & $0.89$ & $1.16$ & $0.94$\\ 
    $\epsilon_{\rm cut}/\mu$ &  $ 1.53 $ & $1.87$ & $2.55$ & $3.12$ \\ \hline \hline
    \end{tabular}
      \caption{System parameters for testing the energy cutoff dependence, for the highest and lowest temperatures used in the SPGPE simulations of quantum vortex decay. The upper and lower cutoffs are $\epsilon^{\uparrow}_{\rm cut}=1.1\epsilon_{\rm cut}$, $\epsilon^{\downarrow}_{\rm cut}=0.9\epsilon_{\rm cut}$, for the values of $\epsilon_{\rm cut}$ given in Table \ref{paramsMod}.}
         \label{paramsCut}
\end{center}
\end{table}
An essential part of SPGPE calculations involves checking for cutoff independence by performing a consistent variation of the energy cutoff. The cutoff independence of the theory has previously been established for $\gamma$-SPGPE, and it must be reassessed for the $\ve$-SPGPE terms. For the lowest and highest temperatures considered in this work, we vary the cutoff and consistently change all parameters to maintain the same total atom number at the temperature of interest. This results in a set of parameters for testing cutoff dependence shown in Table \ref{paramsCut}. The values of $\bar t$ resulting from increasing and decreasing $\ec$, at constant $T$, $N$ are given in \fref{fig:lifetime}. The ensemble-average lifetime is little-changed by consistent variation of $\ec$; indeed, the values of $\bar t$ resulting from our cutoff variation are generally within the ensemble standard deviation of the mean. 
\section{Comparison of theories at $T=0.6T_c$}
The marriage of the Beliaev symmetry breaking approach with kinetic theory in the context of finite-temperature BEC~\cite{Zaremba:1999iu} was a significant development both for fundamental understanding~\cite{QuantumGases,GriffinChapter,Wright:2013ud}, and for modelling of trapped gas experiments~\cite{Proukakis:2008eo,Allen:2013wy}. The resulting `ZNG' theory has proven an accurate description of the essential finite temperature physics far from the region of critical fluctuations.
A prime example is offered by the decay of collective modes~\cite{Jackson:2001eg,Jackson:2002bu,Jackson:2002js,Jackson:2003hd} where ZNG gives a reliable account of the experimental observations; recent work of relevance in our present context has focused on vortex dynamics~\cite{Jackson:2009jo,Allen:2013cs}. For the purpose of the current discussion, 
the ZNG approach contains 
a fully dynamical description of the condensate, the non-condensate, and their particle-exchanging interactions. This property underpins the accuracy of ZNG for low temperatures, and near equilibrium dynamics, but also suggests the need for new tests of finite temperature theory.
\par
In the present work our main focus is the vortex escape dynamics in the full SPGPE theory. However, an interesting comparison can be made between the full SPGPE and ZNG  in a regime where both theories satisfy their respective validity criteria. For SPGPE theory the criteria are (i) that for characteristic single particle energy $\hbar\omega$, $\hbar\omega/k_BT\ll1$, i.e. that the system may be considered \emph{high temperature}, (ii) that the cutoff $\ec$ is chosen to give a definition of the $I$- and $C$-regions that is consistent with the truncated Wigner approximation~\cite{Gardiner:2003bk}, and finally (iii) that $\mu/k_BT\ll 1$. For the ZNG theory, as implemented in Refs.~\cite{Jackson:2009jo,Allen:2013cs}, the system must be cold enough to give a highly occupied condensate mode, justifying a spontaneous symmetry breaking approach to defining the mean-field order parameter~\cite{Zaremba:1999iu}. For our chosen parameters, the characteristic trap frequency is $\bar\omega=2\pi\times 238 {\rm s}^{-1}$. At temperature $T/T_c=0.6$, we then have $\hbar\bar\omega/k_BT=0.041$, while the ideal gas condensate fraction is $N_0/N=0.78$. The final criteria for our system reads $\mu/k_BT=0.63$. However, the approximation underlying this criterion is not fundamental to the SPGPE derivation~\cite{Gardiner:2003bk}; we accept that some small error may arise as a result of only weakly satisfying the inequality, and use this parameter choice for comparison of several theoretical approaches to finite temperature BEC dynamics. 
\par
A fundamental difference in the initial conditions occurs in setting up the comparison. As a central vortex is a metastable excited state of the BEC, there must be a source of noise to induce symmetry breaking and initiate the dissipative decay. The SPGPE contains thermal noise allowing the study of the previous section to start from $r_v(0)=0$. In this section we compare a range of approaches, including several that have no source of initial symmetry breaking. Thus for numerical convenience we initialise the vortex at an appreciable radius $r_v(0)=r_0 $.
\fref{fig1b} shows the radial position of the vortex as a function of time, comparing several theoretical approaches to modelling the vortex decay. The ensemble averaged radius found from the full SPGPE simulations grows much more rapidly than the $\gamma$-SPGPE. The latter result also coincides closely with that computed from our ZNG simulation. The $\ve$-SPGPE result is not shown as the ensemble average is indistinguishable from that of the full SPGPE. Single trajectories from the full SPGPE and $\gamma$-SPGPE are also shown, choosing representative trajectories for which $r_v(t)$ remains fairly close to $\langle r_v(t)\rangle$. 
To clarify the effect of noise we compare the stochastic evolution of $\varepsilon$-SPGPE and $\gam$-SPGPE with corresponding noise-free (deterministic) simulations, referred to as the energy-damped GPE (edGPE), and number-damped GPE (dGPE), respectively. Both deterministic equations generate slower vortex decay than their associated stochastic differential equations, as should be expected on physical grounds. Note that we start the deterministic simulations with the vortex at $r_v(0) =r_0=1.3\mu{\rm m}$, and for ease of comparison we delay the results for each data set by $t_0$ such that $\langle r_v(t_0)\rangle=r_0$ in the corresponding stochastic ensembe average. To construct a ZNG simulation suitable for comparison to the SPGPE, we first find $N_0$ from the SPGPE equilibrium ensemble at $T=0.6T_c$, for our  total atom number $N$. Then we use a combination of $N_0$, $N$, and $T$ as inputs to the ZNG theory to find an initial state on which to phase imprint a vortex. We have found our results to be practically independent of whether we match the total number of atoms, or the condensate atom number within ZNG to the total number or Penrose-Onsager condensate number of the c-field method respectively; this provides a significant additional test of our numerics and the close correspondence between the two methods. In the ZNG simulation the vortex begins its escape from $r_v(0)=r_0=1.33\mu{\rm m}$, and a time delay is again used in the plot to aid comparison with the $\gamma$-SPGPE ensemble average. The ZNG vortex dynamics agrees qualitatively with $\gamma$-SPGPE ensemble average, in the interval where both may be computed. The ensemble average $\langle r_v(t)\rangle$ is not well defined for the $\gamma$-SPGPE once the vortex leaves the detection region ($r_v(t)>r_f$) for any trajectory in the ensemble. However, comparing ZNG with a representative trajectory from $\gamma$-SPGPE, we observe that the two agree semi-quantitatively. 
\par
The close agreement between $\gamma$-SPGPE and ZNG is noteworthy, and requires further comment. The ZNG description is based on a symmetry-breaking decomposition, which ultimately gives rise to an equation for the condensate self-consistently coupled to a quantum Boltzmann equation for the non-condensate. Although the noncondensate treatment involves stochastic sampling of a test particle ensemble to model noncondensate-noncondensate and condensate-noncondensate collisions~\cite{Jackson:2002js}, it does not have an explicit noise term associated with these processes, in contrast to the SPGPE.
All interactions modelled within ZNG involve particle transfer (either to the condensate, or to different regions of the noncondensate). 
This is similar to the $\gamma$-SPGPE term, for which the $C$-field  
contains the condensate \emph{and} many low-energy non-condensate modes,  
thus including all of the inherent collisions and particle transfer between condensate and non-condensate (within the classical field approximation, and up to a cutoff).
Both the simple growth SPGPE, and the ZNG theory explicitly account for thermal-thermal collisions, although their representations of these processes differ in important details; specifically, the simple growth SPGPE neglects thermal-thermal collisions for the population above the cutoff, whereas ZNG includes these processes. However, the contribution of such terms on vortex dynamics has been found to be relatively weak~\cite{Allen:2013wy}, partly justifying the evident similarity of $\gamma$-SPGPE and ZNG found here.
We also note that while the ZNG model gives the condensate density directly, in $\gamma$-SPGPE, a single run of which can be thought of as qualitatively describing a single experimental realisation, the condensate must be found via an ensemble average. 
Finally, we should note that while we have focused on a particular energy-transfer process within the full SPGPE, which is evidently not contained within ZNG, there are other cases where the ZNG method does contain the dominant energy-damping process, thus fully describing the resulting damping: a typical example of this is Landau damping \cite{Jackson:2003hd}. Further discussion of the relation between the two theoretical approaches can be found in Refs.~\cite{Wright:2013ud,GriffinChapter}.
\section{Conclusions}
Dissipative theories of Bose-Einstein condensates play a central role in descriptions of phase transition dynamics and decay of excitations such as vortices and solitons, yet their detailed predictions are largely untested experimentally, and thus far have almost universally focused on the process of particle exchange with the reservoir as the mechanism of dissipation~\cite{Zaremba:1999iu,Stoof:1999tz,Anglin:1999fn,Tsubota2002,Wouters09a}; indeed, this interaction is sufficient to account for processes that are fundamentally associated with condensate growth via evaporative cooling \cite{Stoof:2001wk,Weiler:2008eu}.
Within SPGPE theory, early works argued for the neglect of all the energy-damping (``scattering") processes on the grounds that the terms should be small for quasi-equilibrium systems~\cite{Bradley:2008gq,Rooney:2010dp,Blakie:2008isa}.
In applying the full SPGPE to the decay of a quantum vortex, we find a regime where the damping process associated only with energy exchange dominates  over collisional particle transfer, hastening the escape of the vortex, and providing a clear experimental test of high-temperature theories of superfluid dynamics.  

The most surprising result of our work is that the difference between the full SPGPE and $\gamma$-SPGPE (and also ZNG) increases as the temperature decreases, reaching its most pronounced departure at $T=0.6T_c$, where we are confident in comparing the three theories. One may be tempted to expect a smoother cross-over between predictions of the theories, and indeed this may be evident at lower temperatures not treated in this work. However, for sufficiently low temperatures the SPGPE cutoff cannot be chosen consistently for truncated Wigner validity; furthermore, the linearization of the reservoir interaction eventually becomes invalid~\cite{Gardiner:2003bk}. We believe our predictions suggest an interesting regime for experimental testing. The need for rigorous testing~\cite{Morgan:2003bv} of dissipative quantum field theories describing open systems is further motivated by interest in superfluid internal convection~\cite{Gilz:2011jma,Gilz:2015eq}, far-from equilibrium dynamics~\cite{Polkovnikov:2011iu,Bradley:2014a}, thermalisation~\cite{Rigol:2009ew,Rigol:2008bf}, and critical phenomena~\cite{Damski10a,Su:2013dh,Das:2012ki,Navon:2015jd,McDonald:2015ju}. 

\acknowledgements
SJR acknowledges support from the Victoria University of Wellington during the early stages of this work, and from the University of Otago. ASB is supported by a Rutherford Discovery Fellowship administered by the Royal Society of New Zealand.
NPP would like to acknowledge discussions with Kean Loon Lee, Ian Moss and Eugene Zaremba; NPP/AJA acknowledge funding from EPSRC (grant number EP/I019413/1).

 
%

\end{document}